\documentclass[aps,prl,twocolumn,superscriptaddress]{revtex4-1}
\usepackage{bm}
\usepackage{graphicx}
\usepackage{color}
\usepackage{braket}
\usepackage{amsmath,amssymb,amsfonts,amsthm,mathtools}
\usepackage{enumerate}
\usepackage{enumitem}
\usepackage[colorlinks=true,linkcolor=blue,anchorcolor=red,citecolor=blue,urlcolor=blue]{hyperref}
\usepackage{titlesec}
\usepackage{float}
\usepackage{multirow}
\usepackage[title]{appendix}
\usepackage{pdfpages}
\usepackage{changes}

\newcommand{\vect}[1]{\boldsymbol{#1}}

\makeatletter
\AtBeginDocument{\let\LS@rot\@undefined}
\makeatother

\makeatletter
\renewcommand\NAT@biblabelnum[1]{#1.}

\usepackage{pifont}

\makeatother

\begin{document}

\title{Quantum Geometric Entropy Production and Entropy Hall Effect}
\author{Longjun Xiang}
\affiliation{College of Physics and Optoelectronic Engineering, Shenzhen University, Shenzhen 518060, China}
\author{Jinxiong Jia}
\affiliation{Department of Physics, University of Science and Technology of China, Hefei, Anhui 230026, China}
\author{Jian Wang}
\email{jianwang@hku.hk}
\affiliation{College of Physics and Optoelectronic Engineering, Shenzhen University, Shenzhen 518060, China}
\affiliation{Quantum Science Center of Guangdong-Hongkong-Macao Greater Bay Area, Shenzhen 518045, China}
\affiliation{Department of Physics, The University of Hong Kong, Pokfulam Road, Hong Kong, China}

\begin{abstract}
Quantum geometry has been extensively linked to charge, spin, orbital, and thermal transport in crystalline solids,
but its imprint on entropy transport remains largely unexplored.
Here we develop a quantum theory of entropy transport for Bloch electrons under an electric field.
{Within the relaxation-time approximation,
we show that the leading and subleading contributions to entropy production reproduce the Joule heating relation,
with the former arising from the linear Drude current and the latter from the nonlinear Drude current and the intrinsic nonlinear Ohmic current.
Microscopically, near topological Dirac band crossings and anticrossings,
these contributions are controlled by the quantum metric and the fully symmetrized quantum metric dipole, respectively.
We further predict a Berry-curvature-induced entropy Hall effect,
which, together with the local heat Hall current, provides a simple way to recover the modern theory of orbital magnetization.}
Finally, we derive general relations between entropy and charge currents,
offering a practical route to probe entropy flow experimentally.
\end{abstract}

\maketitle

\noindent \textbf{INTRODUCTION}\\
Entropy production lies at the heart of nonequilibrium thermodynamics~\cite{DeGroot},
as it quantifies the fundamental distinction between reversible and irreversible processes~\cite{RMP2021}.
Determining whether a nonequilibrium transport process is dissipative also requires a direct analysis of entropy production.
Within semiclassical transport theory, such an analysis is usually carried out indirectly by combining the near-equilibrium assumption~\cite{GaoPRR}
with the first law of thermodynamics, $du=\mu dn+Tds$,
where $u$, $n$, and $s$ denote the energy, particle number, and entropy, respectively.
Nevertheless, a fully quantum-mechanical description of entropy current~\cite{Strunk2021} and entropy production,
particularly for Bloch electrons in crystalline solids, has yet to be established.

Meanwhile, quantum geometry~\cite{PRLessay, LuReview, YanReview, Ohm, YuQM, QueirozReview, XuRMP},
such as Berry curvature and quantum metric,
has emerged as a universal framework for understanding transport phenomena in crystalline solids~\cite{Luttinger1954,
Xiao2010, Nagaosa2010, universal, ZhangScience, Xiaospin2025, Xiangspin2025, OrbitalHall, valley, Qin2011, PatrickLee, ZhangLFPRL, XiaoD2006,
FuBCD, GaoY2014PRL, BPT1, BPT2, YanPRL2024, CulcerPRBL, GangSu2022, Jia2024, Ulrich, ShiLK, XuSY2023, Wang2023, Han2024, XiangMap, YanPRR, AhnPRX, DaoDao1},
including Hall effects associated with charge~\cite{FuBCD, GaoY2014PRL, BPT1, BPT2, YanPRL2024, CulcerPRBL, GangSu2022,
Jia2024, Ulrich, ShiLK, XuSY2023, Wang2023, Han2024, XiangMap},
spin~\cite{universal, ZhangScience, Xiaospin2025, Xiangspin2025}, orbital~\cite{OrbitalHall}, and valley~\cite{valley} degrees of freedom.
{Interestingly, Berry curvature can drive an anomalous Ettingshausen effect~\cite{XiaoD2006} for energy transport.}
However, it is unclear whether Berry curvature can likewise induce an entropy Hall effect,
although the first law of thermodynamics places entropy on the same footing as energy.
In addition, how entropy production can be understood in terms of quantum geometry remains less explored.

\begin{figure}[t!]
\includegraphics[width=0.90\columnwidth]{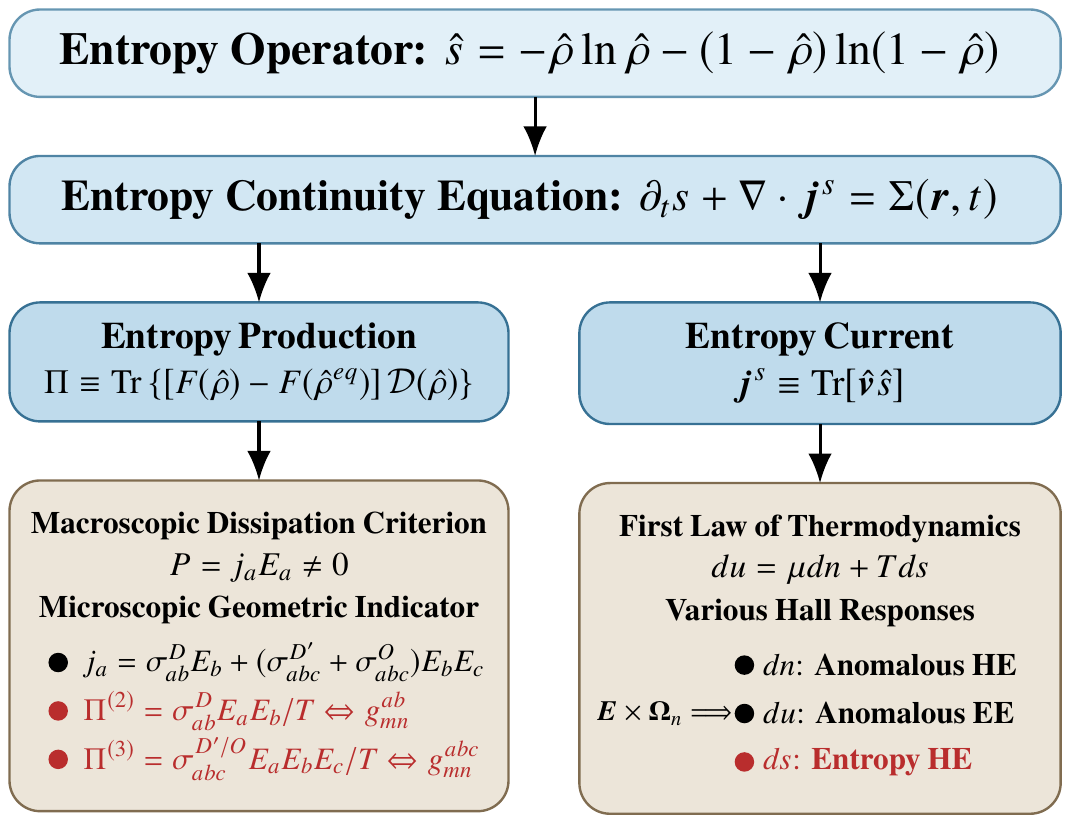}
\caption{{Quantum geometric entropy transport in crystalline solids.
The single-particle entropy operator leads to quantum-mechanical expressions for entropy production and entropy current
through the entropy continuity equation. 
Within the relaxation-time approximation and near topological Dirac band crossings and anticrossings,
the leading and subleading contributions to entropy production are controlled by the quantum metric $g_{mn}^{ab}$
and the fully symmetrized quantum-metric dipole $g_{mn}^{abc}$, respectively.
In addition to the anomalous Hall effect (HE) and the anomalous Ettingshausen effect (EE),
Berry curvature $\bm{\Omega}_n$ also gives rise to an entropy HE.}
}\label{Fig1}
\end{figure}

{In this work, we develop a gauge-invariant quantum theory of entropy transport for Bloch electrons in crystalline solids,
as summarized in Fig.~\ref{Fig1}.
Starting from the single-particle entropy operator of noninteracting fermions,
we derive quantum-mechanical expressions for entropy production and entropy current in the presence of an electric field.
Within the relaxation-time approximation,
we show that the leading and subleading contributions to entropy production both reproduce the macroscopic Joule heating relation,
arising from the linear Drude current and from the nonlinear Drude current~\cite{FuBCD}
and the intrinsic nonlinear Ohmic current~\cite{YanPRL2024,CulcerPRBL,GangSu2022,Jia2024}, respectively.
Microscopically, near topological Dirac band crossings and anticrossings,
the same leading and subleading contributions are controlled by the quantum metric and the fully symmetrized quantum metric dipole,
respectively (see lower-left panel of Fig.~\ref{Fig1}), as illustrated in a tilted Dirac model.
Furthermore, we propose an entropy Hall effect, a transverse entropy flow induced by Berry curvature $\vect{\Omega}_n$,
which we demonstrate in a ferromagnetic Rashba model.
Together with the anomalous Hall effect for charge transport~\cite{Nagaosa2010}
and the anomalous Ettingshausen effect for energy/heat transport~\cite{XiaoD2006},
this establishes that all three extensive quantities entering the first law of thermodynamics can exhibit Hall effects
driven by the anomalous velocity~\cite{Nagaosa2010} $\vect{E} \times \vect{\Omega}_n$ (see lower-right panel of Fig.~\ref{Fig1}).
Remarkably, combining the transport entropy Hall current with the local heat Hall current
allows the orbital magnetization to be extracted directly,
reproducing the modern theory of orbital magnetization~\cite{Xiao2005OM, Thonhauser2005OM, Ceresoli2006OM, Shi2007OM}.
Finally, we derive general relations between entropy and charge currents, 
providing a practical route to experimentally probe entropy flow.}

\bigskip
\noindent \textbf{RESULTS}\\
\textbf{Entropy continuity equation}---The entropy operator $\hat{s}$ for a system of noninteracting fermions
is defined as~\cite{Einstein, Landau, APL}
\begin{align}
\hat{s} = -\hat{\rho} \ln \hat{\rho} - (1-\hat{\rho}) \ln (1-\hat{\rho}).
\end{align}
{Here $\hat{\rho}$ is the single-particle density matrix rather than the full many-body density matrix,
so that $\hat{s}$ is an entropy operator in the single-particle Hilbert space.
Its diagonal element in the position representation defines the entropy density,
$s(\bm r,t)\equiv\langle\bm r|\hat{s}|\bm r\rangle$,
while its trace gives the total entropy of the noninteracting fermionic system, $S=\int d\bm r\,s(\bm r,t)=\mathrm{Tr}[\hat{s}]$.}
Throughout this work, we set $k_B=e=\hbar=1$ unless otherwise specified.

Note that $\hat{\rho}$ obeys the quantum Liouville equation $i\partial_t \hat{\rho} = [\hat{H}, \hat{\rho}]$
and the single-particle entropy operator $\hat{s}$ is a function solely of $\hat{\rho}$, so we have
\begin{align}
i\partial_t \hat{s} = [\hat{H}, \hat{s}],
\label{seq}
\end{align}
where $\hat{H}$ is the system Hamiltonian.
In this work, we consider $\hat{H}=\hat{H}_0+\vect{E}\cdot\hat{\vect{r}}$,
where $\hat{H}_0=\hat{\vect{p}}^2/(2m)+V(\hat{\vect{r}})$ is the Hamiltonian of the crystalline solid and $\vect{E}$ is the electric field.
Taking the time derivative of the entropy density $s(\vect{r}, t)$ and using Eq.~\eqref{seq},
we obtain the entropy continuity equation~\cite{sup}
\begin{align}
\partial_t s(\vect{r}, t) + \nabla_{\vect{r}}\cdot\vect{j}^s(\vect{r}, t) = 0.
\label{scont}
\end{align}
Here $\vect{j}^s(\vect{r}, t) \equiv \langle \vect{r} |\{\hat{\vect{v}},\hat{s}\}|\vect{r}\rangle/2$ is the entropy current density,
where $\hat{\vect{v}}=i[\hat{H}_0, \vect{r}]$ is the velocity operator and $\{,\}$ denotes the anticommutator.
Integrating $\vect{j}^s(\vect{r}, t)$ over $\vect{r}$, we obtain
\begin{align}
\vect{j}^s
=
\dfrac{1}{2} \text{Tr}\left[ \{\hat{\vect{v}}, \hat{s}\} \right]
=
\text{Tr}[\hat{\vect{v}}\hat{s}],
\label{JS}
\end{align}
which gives the quantum-mechanical definition of entropy current.
Equations~\eqref{JS} and~\eqref{seq} allow the entropy current of Bloch electrons
to be evaluated within the same unitary density-matrix formalism used for linear and nonlinear charge currents~\cite{Kraut1979,
Baltz1981,Sipe1995,Sipe2000,Rappe2012,JEMoore2019,Yan2019,Wanghua2020,Juan2020,WatanabePRX}.

Equation~\eqref{scont} shows that the total entropy is conserved under the unitary evolution~\cite{PRXquantum}
because dissipation in solids has been ignored for the moment.
After incorporating dissipation, we replace the quantum Liouville equation by
\begin{align}
\partial_t \hat{\rho} = - i[\hat{H}, \hat{\rho}] + \mathcal{D}(\hat{\rho}),
\label{rhoeq}
\end{align}
where $\mathcal{D}(\hat{\rho})$ is the dissipator accounting for the relevant dissipation processes.
Correspondingly, the entropy continuity equation becomes
\begin{align}
\partial_t s(\vect{r}, t) + \nabla_{\vect{r}} \cdot \vect{j}^s(\vect{r}, t)
=
\Sigma(\vect{r}, t),
\label{scont1}
\end{align}
{where~\cite{sup}
\begin{align}
\Sigma(\vect{r}, t) \equiv \dfrac{1}{2\pi i} \oint_C dz K(z)\langle \vect{r}| \dfrac{1}{z-\hat{\rho}} \mathcal{D}(\hat{\rho}) \dfrac{1}{z-\hat{\rho}}|\vect{r}\rangle
\end{align}
denotes the entropy source determined by the dissipator $\mathcal{D}(\hat{\rho})$.
Here $K(z)=-z\ln z-(1-z)\ln(1-z)$.}

Integrating Eq.~\eqref{scont1} over $\vect{r}$ and using $\int d\vect{r} \nabla_{\vect{r}} \cdot \vect{j}^s(\vect{r}, t)=0$,
we obtain the rate of change of the total entropy~\cite{sup},
$\dot{S} = \int d\vect{r} \Sigma(\vect{r}, t) = \text{Tr}\left[ F(\hat{\rho}) \mathcal{D}(\hat{\rho}) \right]$,
where $\dot{S}=\partial_t S$, with $S$ being the total entropy defined above,
and $F(z) \equiv K'(z)=\ln [(1-z)/z]$.
Moreover, using Spohn's decomposition~\cite{RMP2021} with respect to the equilibrium density matrix $\hat{\rho}^{eq}$,
the rate of change of the total entropy can be expressed as $\dot{S} = \Pi + \Phi$,
where $\Phi \equiv \text{Tr}\left[ F(\hat{\rho}^{eq})\mathcal{D}(\hat{\rho})\right]$ and
\begin{align}
\Pi
\equiv
\text{Tr}\{\left[ F(\hat{\rho})-F(\hat{\rho}^{eq}) \right] \mathcal{D}(\hat{\rho})\}
\label{Pi}
\end{align}
defines the entropy-production rate.
For a physical dissipative process, the second law of thermodynamics requires~\cite{PRXquantum} $\Pi \geq 0$.
{Here, the expressions involving $F(\hat{\rho})$ should first be defined at finite temperature,
where the eigenvalues of the fermionic single-particle density matrix lie in the interval $(0, 1)$,
and the zero-temperature limit should be taken only after the finite-temperature results are obtained.}
Eq.~\eqref{Pi}, together with Eq.~\eqref{rhoeq},
constitutes the quantum-mechanical framework for entropy production by Bloch electrons in crystalline solids,
as discussed below.

\bigskip
\noindent\textbf{Dissipator within the relaxation-time approximation}---To illustrate entropy production,
we adopt the dissipator within the relaxation-time approximation (RTA) by choosing~\cite{Mikhailov, Peres}
\begin{align}
\mathcal{D}(\hat{\rho}) = - \eta \left(\hat{\rho}-\hat{\rho}^{eq}\right),
\label{RTAd}
\end{align}
{where $\eta$ is an infinitesimal positive parameter.
Two remarks are in order.
First, by substituting Eq.~\eqref{RTAd} into Eq.~\eqref{Pi}, we have
\begin{align}
\Pi
=
-\text{Tr}\{\eta\left[ F(\hat{\rho})-F(\hat{\rho}^{eq}) \right]\left( \hat{\rho} - \hat{\rho}^{eq} \right) \} \geq 0.
\label{Pieta}
\end{align}
This follows because $\hat\rho$ and $\hat\rho^{eq}$ are fermionic single-particle density matrices with spectra in $(0,1)$,
while $F(x)=\ln[(1-x)/x]$ is operator-monotone decreasing on this interval. Second, in the perturbative calculations below,
this RTA dissipator is implemented order by order with different infinitesimal parameters for the diagonal and off-diagonal density-matrix elements.
Accordingly, these infinitesimal parameters enter all three factors in Eq.~\eqref{Pieta}:
explicitly through $\eta$, and implicitly through both $F(\hat{\rho})-F(\hat{\rho}^{eq})$ and $\hat{\rho}-\hat{\rho}^{eq}$,
since $\hat{\rho}$ itself depends on $\eta$ through Eq.~\eqref{rhoeq}.
We retain these infinitesimal parameters until the complete entropy production is assembled.
In the limit $\eta \rightarrow 0^+$,
the remaining divergent factors are regularized in terms of a finite relaxation time $\tau$~\cite{Jia2024},
yielding the entropy production from the Drude currents,
while the finite regular part gives that from the intrinsic nonlinear Ohmic current~\cite{YanPRL2024,CulcerPRBL,GangSu2022,Jia2024}.
We remark that this perturbative RTA prescription reproduces the charge currents obtained from the semiclassical theory,
as shown in the Supplemental Material~\cite{sup}.
In addition, the leading entropy production is shown explicitly below to remain nonnegative,
ensuring the positivity of the total entropy production in the weak-field regime.}

Inserting Eq.~\eqref{RTAd} into Eq.~\eqref{rhoeq} and 
using the eigenbasis of the Bloch Hamiltonian $\hat{\mathcal{H}}_0 = e^{-i\vect{k}\cdot\hat{\vect{r}}}\hat{H}_0e^{i\vect{k}\cdot\hat{\vect{r}}}$
with $\hat{\mathcal{H}}_0|u_n\rangle=\epsilon_n|u_n\rangle$,
where $\epsilon_n \equiv \epsilon_n(\vect{k})$ is the energy of the $n$th Bloch band,
with $\vect{k}$ the crystal momentum, and $|u_n\rangle \equiv |u_n(\vect{k})\rangle$ is the cell-periodic Bloch wavefunction,
we obtain
\begin{align}
&i\partial_t \rho_{mn}
=
\epsilon_{mn} \rho_{mn}
+
i D^a_{mn} \rho_{mn} E_a(t)
\nonumber \\
&+
\sum_l
\left( r^a_{ml}\rho_{ln}-\rho_{ml} r^a_{ln} \right)
E_a(t)
-
i\eta
\left(
\rho_{mn}
-
\rho^{eq}_{mn}
\right).
\label{rhomn}
\end{align}
Here $\rho_{mn}=\langle u_m|\hat{\rho}|u_n\rangle$;
$\rho_{mn}^{eq}=\langle u_m|\hat{\rho}^{eq}|u_n\rangle=\delta_{mn}f_n$,
where $f_n$ is the equilibrium Fermi-Dirac distribution function;
$\epsilon_{mn}=\epsilon_m-\epsilon_n$;
$D^a_{mn}=\partial_a-i(\mathcal{A}_{mm}^a-\mathcal{A}_{nn}^a)$ is the $U(1)$ covariant derivative,
with $\partial_a=\partial/\partial k_a$ and $\mathcal{A}^a_{mm}=\langle u_m|i\partial_a|u_m\rangle$ being the intraband Berry connection;
$r^a_{ml}=\langle u_m|i\partial_a|u_l\rangle$, with $m \neq l$, is the interband Berry connection;
$E_a(t)=E_a(\omega_\alpha)e^{-i \omega_\alpha t}+c.c.$, with $\omega_\alpha = \pm \omega$.
{Equation~\eqref{rhomn} can be solved iteratively.
Expanding $\rho_{mn}=\sum_{j \geq 0} \rho_{mn}^{(j)}$,
where $\rho_{mn}^{(j)} \propto (E_a)^j$ and $\rho_{mn}^{(0)}=\rho_{mn}^{eq}$,
we find
\begin{align}
\partial_t \rho_{mn}^{(j)}
&
+
\left[ i\epsilon_{mn} + \eta^{(j)}_{mn} \right] \rho_{mn}^{(j)}
=
D^a_{mn} \rho_{mn}^{(j-1)}
E_a(t)
\nonumber \\
&-
i\sum_l
\left[ r^a_{ml}\rho_{ln}^{(j-1)}-\rho_{ml}^{(j-1)} r^a_{ln} \right]
E_a(t),
\label{rhomn1}
\end{align}
where $\eta_{mn}^{(j)} \equiv [1+(j-1)\delta_{mn}]\xi$, with $\xi \rightarrow 0^+$,
is the infinitesimal parameter associated with the $j$th-order density-matrix element $\rho_{mn}^{(j)}$.
For the diagonal and off-diagonal components of $\rho_{mn}^{(j)}$,
we have $\eta_{mm}^{(j)}=j\xi$ and $\eta_{mn}^{(j)}=\xi$ for $m \neq n$, respectively.
Throughout this work, we adopt the Einstein summation convention over repeated indices $a$, $b$, and $c$.}

{
Solving Eq.~\eqref{rhomn1} for $j=1$ and taking the DC limit ($\omega_\alpha \rightarrow 0$),
we obtain~\cite{sup}
\begin{align}
\rho_{mn}^{(1)}
&=
\dfrac{\delta_{mn} \partial_a f_n}{\xi}E_a + \dfrac{f_{mn}r^a_{mn}}{\epsilon_{mn}}E_a,
\label{rhomn1DC}
\end{align}
where $f_{mn}=f_m-f_n$.
The first term of Eq.~\eqref{rhomn1DC},
after regularizing the divergent factor $1/\xi$ by the replacement $1/\xi \rightarrow \tau$,
gives the linear Drude current,
while its second (regular) term gives the intrinsic anomalous Hall current~\cite{Nagaosa2010},
consistent with the semiclassical theory within the RTA.
Similarly, solving Eq.~\eqref{rhomn1} for $j=2$ and taking the DC limit, we obtain~\cite{sup}
\begin{align}
\rho_{mn}^{(2)} &= \dfrac{1}{\xi^2}\dfrac{\delta_{mn}\partial_{a}\partial_bf_m}{2}E_aE_b + \dfrac{1}{\xi}\dfrac{r^a_{mn}\partial_bf_{mn}}{\epsilon_{mn}}E_aE_b
\nonumber \\
&+\dfrac{ir^a_{mn}\partial_bf_{mn}}{\epsilon_{mn}^2}E_aE_b-\sum_{l} \dfrac{\delta_{mn}f_{ml}g^{ab}_{ml}}{\epsilon_{ml}^2}E_aE_b
\nonumber \\
&+\dfrac{1}{i\epsilon_{mn}}D^a_{mn}\left( \dfrac{f_{mn}r^b_{mn}}{\epsilon_{mn}}\right)E_aE_b
\nonumber \\
&-\sum_{l}
\dfrac{1}{\epsilon_{mn}}\left( \dfrac{r^a_{ml}r^b_{ln}f_{ln}}{\epsilon_{ln}} - \dfrac{f_{ml}r^b_{ml}r^a_{ln}}{\epsilon_{ml}} \right)E_aE_b.
\label{rhomn2DC}
\end{align}
Here the first and second terms of Eq.~\eqref{rhomn2DC},
after regularizing the divergent factors by the replacements $1/\xi^2 \rightarrow \tau^2$ and $1/\xi \rightarrow \tau$,
respectively, give the nonlinear Drude current and the extrinsic nonlinear Hall current~\cite{FuBCD},
while the remaining regular terms give the intrinsic nonlinear Hall~\cite{GaoY2014PRL}
and Ohmic~\cite{YanPRL2024,CulcerPRBL,GangSu2022,Jia2024} currents,
consistent with the semiclassical theory within the RTA~\cite{sup}.
With Eqs.~\eqref{rhomn1DC} and \eqref{rhomn2DC},
we are ready to evaluate the leading and subleading contributions to entropy production.}

\begin{figure*}[t!]
\includegraphics[width=1.95\columnwidth]{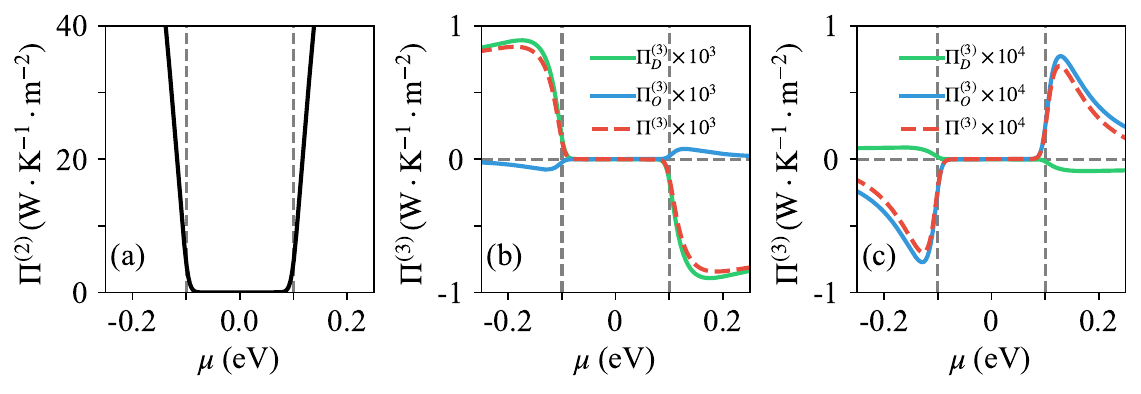}
\caption{{Entropy production in the tilted Dirac model
$\hat{\mathcal{H}}_0=t_xk_x+ v_xk_x\hat{\sigma}_x+v_yk_y\hat{\sigma}_y+\Delta\hat{\sigma}_z$.
(a) Leading entropy production for $\tau=10^{-14}\,\mathrm{s}$.
(b) and (c) Nonlinear Drude contribution $\Pi_D^{(3)}$,
intrinsic nonlinear Ohmic contribution $\Pi_O^{(3)}$, and their sum $\Pi^{(3)}=\Pi^{(3)}_D+\Pi^{(3)}_O$
for $\tau=10^{-14}\,\mathrm{s}$ and $10^{-15}\,\mathrm{s}$, respectively.
Parameters are $t_x=0.2\mathrm{eV}\cdot\mathrm{\AA}$,
$v_x=v_y=1\mathrm{eV}\cdot\mathrm{\AA}$, $\Delta=0.1\,\mathrm{eV}$, $T=50\,\mathrm{K}$, and $E_x=10^4\,\mathrm{V/m}$.
The vertical dashed lines mark the band edges.}
}
\label{FIG2}
\end{figure*}

\bigskip
\noindent\textbf{Quantum geometric entropy production}---{To begin with,
we define $A=\hat{\rho}^{eq}$, $B=\hat{\rho}-\hat{\rho}^{eq}=\sum_{j \geq 1} \hat{\rho}^{(j)}$, where $\hat{\rho}^{(j)} \propto (E_a)^j$,
and introduce $g(t; A, B)=F(A+tB)$, where $F(z)=\ln[(1-z)/z]$ and $t$ is a real parameter.
Expanding $g(t;A, B)$ around $t=0$ in terms of Fr\'echet derivatives,
we have $g(t; A, B) = \sum_{i \geq 0} g^{(i)}(0; A, B) t^i/i!$,
where $g^{(i)}(0; A, B)=d^i g(t; A, B)/dt^i|_{t=0}$,
and therefore $F(\hat{\rho})=g(1;A,B)=\sum_{i \geq 0} g^{(i)}(0; A, B)/i!$.
Within the perturbative RTA, the dissipator is written as
$\mathcal{D}_{mn}(\hat{\rho})=-\sum_{j \geq 1} \eta_{mn}^{(j)}\rho_{mn}^{(j)}$.
Substituting this dissipator and the Fr\'echet expansion into Eq.~\eqref{Pi},
and using $g^{(0)}(0; A, B)=F(\hat{\rho}^{eq})$,
the entropy production in the eigenbasis of $\hat{\mathcal{H}}_0$ becomes~\cite{sup}
\begin{align}
\Pi 
= -\sum_{\substack{
i,j,j_1,\ldots,j_i\geq1\\
m,n,l_1,\ldots,l_{i-1}
}}
& \int_k \eta_{nm}^{(j)} F^{[i]}(f_m, f_{l_1}, \cdots, f_{l_{i-1}}, f_n)
\nonumber \\
& \times
\rho_{ml_1}^{(j_1)}
\rho_{l_1l_2}^{(j_2)}
\cdots
\rho_{l_{i-1}n}^{(j_i)}
\rho_{nm}^{(j)},
\label{PiRTAexpand}
\end{align}
where each term scales as $(E_a)^q$, with $q=j_1+j_2+\cdots+j_i+j$, and
\begin{align}
F^{[i]} (f_m,\cdots,f_n)
=
\oint_C
\dfrac{dz}{2 \pi i}
\dfrac{F(z)}{(z-f_m) \cdots (z-f_n)}.
\end{align}}

{At the leading order in the electric field,
$\Pi^{(2)} \propto E_aE_b$, only $i=1$ and $(j_1,j)=(1,1)$ contribute.
Using $F^{[1]}(f_m, f_m) = -1/[f_m(1-f_m)]$ and $F^{[1]}(f_m, f_n) = \epsilon_{mn}/(Tf_{mn})$ for $m \neq n$,
where $T$ is the temperature, we obtain~\cite{sup}
\begin{align}
\Pi^{(2)}
&=
\xi\sum_m\int_k
\left[
\frac{|\rho^{(1)}_{mm}|^2}
{f_m(1-f_m)}
-
\sum_{n \neq m}
\frac{\epsilon_{mn}|\rho^{(1)}_{mn}|^2}{T f_{mn}}\right],
\label{Pi2}
\end{align}
which is nonnegative since $\epsilon_{mn}/f_{mn}<0$.
In the limit $\xi\rightarrow 0^+$,
the first term survives since $|\rho_{mm}^{(1)}|^2 \propto 1/\xi^2$,
whereas the second term vanishes since the $|\rho_{mn}^{(1)}|^2$ for $m \neq n$ is regular.
Using $\rho_{mm}^{(1)}$ in Eq.~\eqref{rhomn1DC} and $\partial_a f_m/[ f_{m}(1-f_m)] = -v_m^a/T$,
where $v_m^a=\partial_a \epsilon_m$ is the group velocity, we further obtain~\cite{sup}
\begin{align}
\Pi^{(2)} 
=
-
\sum_{m} \int_k 
\dfrac{\tau v_m^a \partial_b f_m}{T}
E_aE_b
=
\dfrac{j_a^DE_a}{T},
\label{Pi3}
\end{align}
where $j_a^D \equiv -\tau \sum_m \int_k v_m^a \partial_b f_m E_b$ is the linear Drude current.
Therefore, the quantum-mechanical calculation reproduces the macroscopic Joule-heating criterion for dissipation,
$T\Pi^{(2)}=P=\vect{j}\cdot\vect{E}$~\cite{Nagaosa2010}.
Microscopically, using $\partial_b v_m^a=w_{mm}^{ab} + \sum_{n \neq m} 2 \epsilon_{mn}g^{ab}_{mn}$~\cite{NagaosaDrude, XiangMap},
where $w_{mm}^{ab} \equiv \langle u_m|\partial^2_{ab}\hat{\mathcal{H}}_0|u_m\rangle$
and $g^{ab}_{mn}=\text{Re}[r^a_{mn}r^b_{nm}]$ is the (local) quantum metric,
Eq.~\eqref{Pi3} becomes
\begin{align}
\Pi^{(2)}
&=
\sum_{m} \int_k
\dfrac{\tau f_m}{T}
\left[ w^{ab}_{mm} + 2\sum_{n \neq m} \epsilon_{mn}g^{ab}_{mn} \right]
E_a E_b.
\end{align}
Near topological Dirac band crossings and anticrossings,
where $w_{mm}^{ab}$ usually vanishes or becomes negligible~\cite{Kang2025QGT,Kim2025QMT},
the leading entropy production is then controlled by the quantum metric.}

{
We next evaluate the subleading entropy production $\Pi^{(3)}\propto E_aE_bE_c$.
According to Eq.~\eqref{PiRTAexpand},
there are two types of third-order contributions: $i=1$ with $(j_1,j)=(1,2),(2,1)$,
and $i=2$ with $(j_1,j_2,j)=(1,1,1)$.
We therefore write $\Pi^{(3)} = \Pi^{(3;1)}+\Pi^{(3;2)}$,
where
\begin{align}
\Pi^{(3;1)}
=&
-\sum_{mn}\int_k
\eta_{nm}^{(2)}
F^{[1]}(f_m,f_n)
\rho_{mn}^{(1)}\rho_{nm}^{(2)}
\nonumber \\
&-
\sum_{mn}\int_k
\eta_{nm}^{(1)}
F^{[1]}(f_m,f_n)
\rho_{mn}^{(2)}\rho_{nm}^{(1)}
,
\label{eq:Pi31}
\\
\Pi^{(3;2)}
=&
-\sum_{mnl}\int_k
\eta_{nm}^{(1)}
F^{[2]}(f_m,f_l,f_n)
\rho_{ml}^{(1)}
\rho_{ln}^{(1)}
\rho_{nm}^{(1)}.
\label{eq:Pi32}
\end{align}
Inserting Eqs.~\eqref{rhomn1DC} and~\eqref{rhomn2DC} into these expressions,
a detailed calculation also reproduces the macroscopic Joule-heating criterion for dissipation~\cite{sup}
\begin{align}
\Pi^{(3)}
&=
\dfrac{j_a^{D'}E_a}{T}
+
\dfrac{j_a^{O}E_a}{T}
=
\Pi^{(3)}_D+\Pi^{(3)}_O,
\label{PIsublead}
\end{align}
Here $j_a^{D'}$ is the nonlinear Drude current~\cite{FuBCD},
while $j_a^O$ is the intrinsic nonlinear Ohmic current~\cite{YanPRL2024, CulcerPRBL, GangSu2022, Jia2024}.
Their expressions can be found in the Methods section.
Thus, in addition to the nonlinear Drude current,
the intrinsic nonlinear Ohmic current contributes to entropy production and is therefore dissipative.
Microscopically, denoting $j_a^{D'}=\sigma_{abc}^{D'}E_bE_c$ and $j_a^O=\sigma_{abc}^OE_bE_c$,
we find that near topological Dirac band crossings and anticrossings,
both $\sigma_{abc}^{D'}$ and $\sigma_{abc}^O$ are controlled by the fully symmetrized quantum-metric dipole
$g^{abc}_{mn} \equiv v_m^a g^{bc}_{mn} + v_m^b g^{ca}_{mn}+v_m^c g^{ab}_{mn}$ (see Methods).
}

{We emphasize that these quantum-geometric quantities determine the symmetry of the conductivity tensors entering entropy production,
but do not fix its magnitude.
The full response also contains additional band-energy and Fermi-surface weighting factors,
so quantum-metric hot spots near band crossings and anticrossings need not coincide with maxima of entropy production.
In addition, although $\Pi^{(3)}$ contributes to dissipation,
it is odd under electric-field reversal and therefore need not be nonnegative by itself.
This does not conflict with the second law of thermodynamics.
In the weak-field regime, the total entropy production is dominated by the positive quadratic contribution $\Pi^{(2)}$,
which guarantees $\Pi=\Pi^{(2)}+\Pi^{(3)}+\cdots \geq 0$.}

{To illustrate,
we consider the tilted Dirac model~\cite{BPT2} $\hat{\mathcal{H}}_0=t_xk_x+v_xk_x\hat{\sigma}_x+v_yk_y\hat{\sigma}_y+\Delta\hat{\sigma}_z$.
As shown in Fig.~\ref{FIG2}(a), the leading entropy production $\Pi^{(2)}$ is positive and is strongly enhanced
when the chemical potential $\mu$ enters the conduction or valence band,
while it is strongly suppressed inside the gap because the Drude current is a Fermi-surface response.
For the same parameters as in Fig.~\ref{FIG2}(a),
the subleading entropy production $\Pi^{(3)}$ shown in Fig.~\ref{FIG2}(b)
is approximately five orders of magnitude smaller than the leading entropy production $\Pi^{(2)}$,
consistent with the positivity of the total entropy production in the weak-field regime.
Although $\sigma_{abc}^{D'}$ and $\sigma_{abc}^{O}$ share a common quantum-metric-dipole origin,
they differ by an overall sign.
Moreover, $\sigma_{abc}^{D'}\propto\tau^2$, whereas $\sigma_{abc}^{O}$ is independent of $\tau$.
Consequently, $\Pi^{(3)}=\Pi_D^{(3)}+\Pi_O^{(3)}$ can reverse sign as $\tau$ is varied,
as demonstrated in Figs.~\ref{FIG2}(b) and \ref{FIG2}(c).
After isolating the subleading entropy production from the leading one via $[\Pi(\vect{E})-\Pi(-\vect{E})]/2=\Pi^{(3)}+\mathcal{O}(E^5)$,
such a sign reversal provides a direct diagnostic for the intrinsic nonlinear Ohmic current~\cite{YanPRL2024, CulcerPRBL, GangSu2022, Jia2024}.}

{We notice that the antisymmetric Berry curvature,
defined by $\Omega^{ab}_{mn}=-2\text{Im}[r^a_{mn}r^b_{nm}]$,
gives rise to transverse Hall currents through the anomalous velocity but cannot appear in the entropy production because $\Omega^{ab}_{mn}E_a E_b=0$.
As a result, both the intrinsic (independent of $\tau$) anomalous Hall current induced by Berry curvature~\cite{Nagaosa2010}
and the extrinsic (proportional to $\tau$) nonlinear Hall current induced by Berry curvature dipole~\cite{FuBCD} are dissipationless.
We emphasize, however, that transverse transport alone does not guarantee the absence of dissipation:
a symmetric off-diagonal conductivity can also generate a transverse current for a particular electric-field direction,
while whether it contributes to entropy production is determined by the full contraction $j_aE_a$.
We further note that the dissipative Drude currents also depend on $\tau$;
therefore, the $\tau$ dependence alone is not a reliable indicator of dissipation~\cite{Xiang2025}.
We also remark that the intrinsic nonlinear Hall effect~\cite{GaoY2014PRL, BPT1, BPT2}
from the field-induced anomalous velocity also originates from the quantum-metric dipole
but is controlled by the combination $\bar{g}^{abc}_{mn}=2v_m^ag^{bc}_{mn}-v_m^bg^{ca}_{mn}-v_m^cg_{mn}^{ab}$ (see Methods),
which does not contribute to entropy production due to $\bar{g}^{abc}_{mn}E_aE_bE_c=0$ and is dissipationless,
distinct from the dissipative intrinsic nonlinear Ohmic current~\cite{YanPRL2024, CulcerPRBL, GangSu2022, Jia2024}.
Finally, we stress that our results were derived for Bloch electrons with an RTA dissipator.
Whether the quantum-metric and quantum-metric-dipole structures of the leading and subleading entropy production,
respectively, remain valid for more general dissipators requires further investigation.}

\begin{figure}[t!]
\includegraphics[width=0.90\columnwidth]{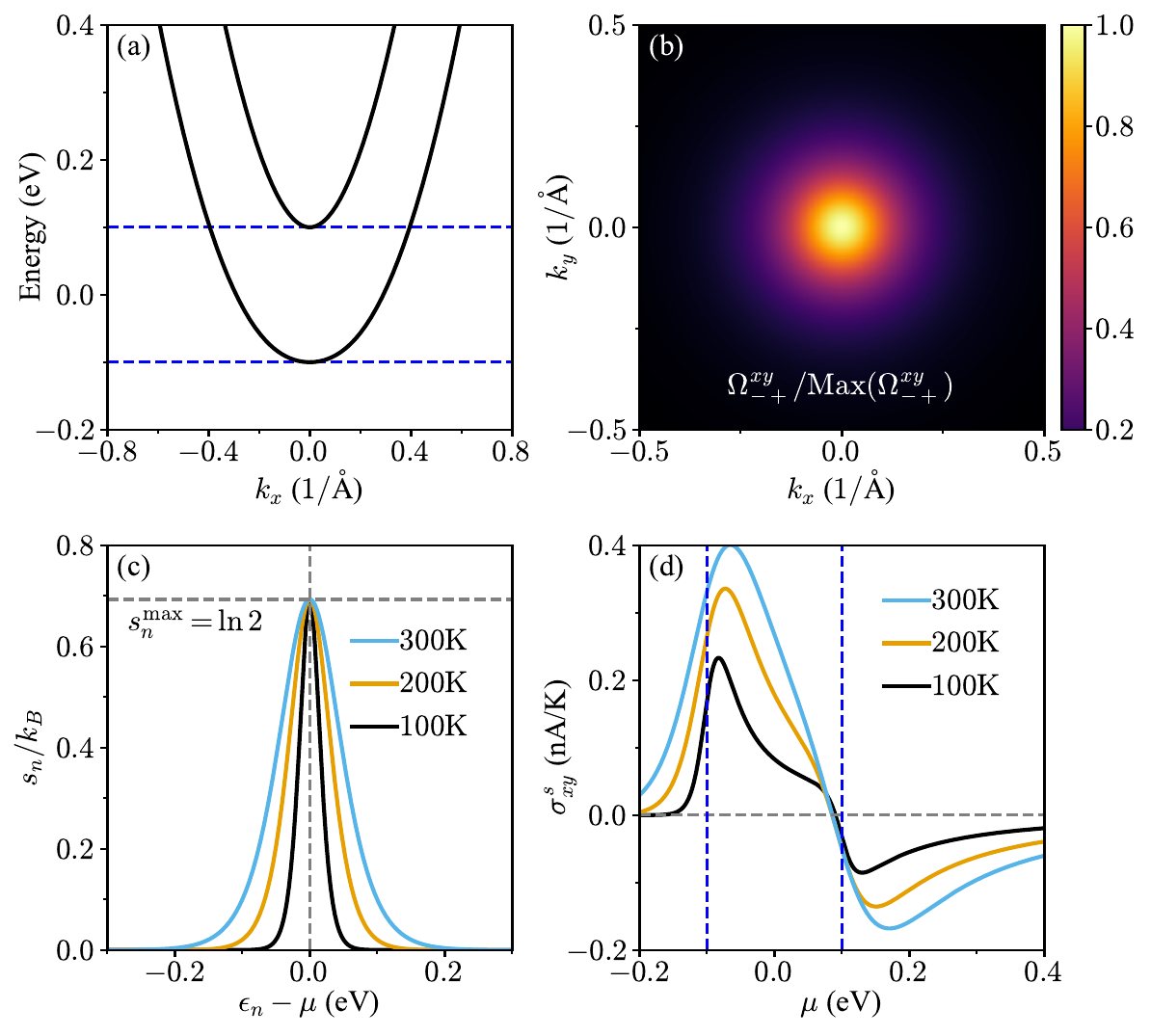}
\caption{{Entropy Hall effect in the ferromagnetic Rashba model~\cite{Xiao2010}
$\hat{\mathcal{H}}_0=k^2/2m+\lambda_R(k_y\hat{\sigma}_x-k_x\hat{\sigma}_y)+\Delta\hat{\sigma}_z$,
where $k^2=k_x^2+k_y^2$.
(a) Band dispersions, where the vertical dashed lines mark the band edges.
(b) Normalized Berry curvature of the lower band.
(c) Entropy weight $s_n$ as a function of $\epsilon_n-\mu$ at different temperatures,
peaked at $\epsilon_n=\mu$ with $s_n^{\rm max}=\ln 2$.
(d) Entropy Hall conductivity $\sigma_{xy}^s$ as a function of $\mu$ at different temperatures.
Parameters are $\lambda_R=0.6\,\mathrm{eV}\cdot\mathrm{\AA}$, $m=1.1m_e$, and $\Delta=0.1\,\mathrm{eV}$.}
}
\label{FIG3}
\end{figure}

\bigskip
\noindent\textbf{Entropy Hall effect}---We next investigate the entropy Hall effect of Bloch electrons.
Since the entropy operator $\hat{s}$ and the density matrix $\hat{\rho}$ obey the same quantum Liouville equation,
while the entropy current $j_a^s=\text{Tr}[\hat{v}^a\hat{s}]$ and charge current $j_a^e=-\text{Tr}[\hat{v}_a\hat{\rho}]$
share the same operator structure,
the entropy current can be obtained from the corresponding charge current through the replacement $f_m\rightarrow -s_m$
within the unitary density-matrix response theory~\cite{Jia2024}.
Here, $s_m=-f_m\ln f_m-(1-f_m)\ln(1-f_m)$ is the band entropy.
Applying this correspondence to the intrinsic anomalous Hall current $j_a^e=-\sum_{mn}\int_kf_m\Omega_{mn}^{ab}E_b$,
we obtain
\begin{align}
j_a^s=\sigma_{ab}^sE_b,
\qquad
\sigma^s_{ab} = \sum_{mn} \int_k s_m \Omega^{ab}_{mn}.
\label{sigmaab}
\end{align}
{An immediate implication of Eq.~\eqref{sigmaab} is
that $T\sigma^s_{ab}$ gives the anomalous Ettingshausen conductivity~\cite{XiaoD2006},
since the transport heat and entropy Hall currents satisfy $j_a^q=Tj_a^s$ in the near-equilibrium linear-response regime~\cite{XiaoC2020}.
By Onsager reciprocity~\cite{Onsager1, Onsager2, Onsager3},
this response is reciprocal to the anomalous Nernst effect~\cite{XiaoD2006},
in which a temperature gradient drives a transverse charge current.
Thus, the three extensive quantities entering the first law of thermodynamics, $du=\mu dn+Tds$,
can all exhibit Berry-curvature-induced Hall responses:
the anomalous Hall effect for particle-number (charge) transport, the anomalous Ettingshausen effect for energy/heat transport,
and the entropy Hall effect for entropy transport.}

{We emphasize that,
for the steady and spatially uniform response considered here,
the entropy Hall current in Eq.~\eqref{sigmaab} obtained from the local definition $j_a^s=\mathrm{Tr}[\hat{v}^a\hat{s}]$
already corresponds to the transport entropy current.
In stark contrast, the transport heat Hall current $j_a^q$ cannot be obtained directly from the local heat-current definition
\begin{equation}
j_a^h=\mathrm{Tr}
\left[\hat v^{a,\hat H_0}\hat\rho\right],
\qquad
\hat v^{a,\hat H_0}
=
\frac{\{ \hat v^a, \hat H_0-\mu\}}{2},
\end{equation}
because the local heat current also contains the circulating magnetization current $j_a^m=-\epsilon_{abc}E_bM^c$, namely, $j_a^q=j_a^h-j_a^m$.
A direct microscopic evaluation of the orbital magnetization $M^c$ in crystalline solids is notoriously difficult
and has long posed a fundamental challenge in solid-state physics~\cite{Xiao2005OM, Thonhauser2005OM, Ceresoli2006OM, Shi2007OM}.
Equation~\eqref{sigmaab} therefore provides an alternative route:
in the near-equilibrium linear-response regime, the transport heat current follows directly from $j_a^q=Tj_a^s$,
without requiring an independent calculation of the orbital magnetization.
Interestingly, since both the transport entropy current and the local heat current can be readily evaluated within the unitary density-matrix response theory,
the difference $j_a^h-Tj_a^s$ can in turn be used to determine the otherwise difficult orbital magnetization.
Specifically, the local intrinsic heat Hall current is given by~\cite{sup}
$j_a^h = \epsilon_{abc}\sum_m\int_k \left[ (\epsilon_m-\mu)\Omega_m^c-m_m^c \right]f_m E_b$.
Subtracting $j^q_a=Tj_a^s=\epsilon_{abc}\sum_m\int_k T s_m\Omega_m^c E_b$ from $j_a^h$,
where $j_a^s$ is obtained from Eq.~\eqref{sigmaab},
we find
$j_a^m = j_a^h-Tj_a^s = \epsilon_{abc}\sum_m\int_k \left( g_m\Omega_m^c-f_m m_m^c \right)E_b = -\epsilon_{abc}E_bM^c$,
which directly yields
\begin{equation}
M^c=
\sum_m\int_k
\left(
f_m m_m^c-g_m\Omega_m^c
\right),
\label{orbitalM}
\end{equation}
where $g_m=(\epsilon_m-\mu)f_m-Ts_m=T\ln(1-f_m)$ is the grand potential.
Eq.~\eqref{orbitalM} reproduces the modern theory of orbital magnetization~\cite{Xiao2005OM, Thonhauser2005OM, Ceresoli2006OM, Shi2007OM}.}

{Since the Berry-curvature-induced entropy Hall effect requires broken time-reversal symmetry,
we illustrate it using the ferromagnetic Rashba model~\cite{Xiao2010},
$\hat{\mathcal{H}}_0=k^2/(2m)+\lambda_R(k_y\hat{\sigma}_x-k_x\hat{\sigma}_y)+\Delta\hat{\sigma}_z$,
where the exchange term $\Delta\hat{\sigma}_z$ breaks time-reversal symmetry and generates a finite Berry curvature.
As shown in Fig.~\ref{FIG3}(a), the Rashba bands are split and gapped near $k=0$.
The corresponding Berry curvature is strongly concentrated near the avoided crossing, as illustrated in Fig.~\ref{FIG3}(b).
According to Eq.~\eqref{sigmaab}, the entropy Hall conductivity is obtained by weighting the Berry curvature with the band entropy $s_n$,
which is sharply peaked at $\epsilon_n=\mu$ with $s_n^{\mathrm{max}}=\ln 2$ [Fig.~3(c)].
Consequently, $\sigma_{xy}^s$ predominantly probes the Berry curvature within a thermal window around the chemical potential
and exhibits pronounced peaks when this window overlaps the Berry-curvature hot spots near the band edges [Fig.~3(d)].
Its sign reversal reflects the opposite Berry curvatures of the two Rashba bands,
while increasing temperature broadens the thermal window and hence the entropy Hall response.}

\bigskip
\noindent\textbf{Measurement of entropy current}---We now discuss possible routes to probe the entropy current experimentally,
which is generally difficult to measure directly~\cite{measurement0, measurement1, measurement2}.
{In the near-equilibrium linear-response regime,
the entropy current can be inferred from heat-current measurements because the heat and entropy currents satisfy $j_a^q=Tj_a^s$~\cite{XiaoC2020}.
This relation also provides a practical route to probe the entropy Hall effect discussed above.
In particular, in a Hall-bar geometry,
an electric field applied along $x$ generates a transverse heat current $j_y^q=\sigma_{yx}^qE_x$,
corresponding to the anomalous Ettingshausen effect,
which in turn corresponds to a transverse entropy current $j_y^s=j_y^q/T$.
Alternatively, by Onsager reciprocity,
the entropy Hall response can be inferred from the anomalous Nernst effect,
in which a longitudinal temperature gradient generates a transverse Hall voltage.
In addition, the linear optical entropy current may be inferred from the linear optical charge current
within the two-band approximation (see Methods).}

We emphasize that the relation $j_a^q=Tj_a^s$ used above applies only to the near-equilibrium linear-response regime
and should not be regarded as a general identity between heat and entropy currents.
For example, when a second-order nonlinear Drude response is symmetry allowed,
the nonlinear Drude entropy current contains an additional contribution
that cannot be reconstructed from the nonlinear Drude heat current using $j_a^q=Tj_a^s$~\cite{sup}.
Therefore, nonlinear entropy currents generally cannot be inferred from heat-current measurements alone.
To address this difficulty, we establish two relations between entropy and charge currents,
which provide practical approaches to infer entropy currents from charge-transport measurements.

First, using the correspondence $f_m \rightarrow -s_m$ established above,
together with $f_m=-\partial_\mu g_m$ and $s_m=-\partial_T g_m$~\cite{XCPRL2020},
where $g_m=-T\ln[e^{-(\epsilon_m-\mu)/T}+1]$ is the grand potential,
we obtain the Maxwell-type relation~\cite{sup}
\begin{align}
\partial_T j^e_a=-\partial_\mu j^s_a,
\end{align}
which holds for both AC and DC driving, includes both intrinsic and extrinsic contributions,
and applies in both linear and nonlinear response regimes.

Second, in the low-temperature regime where the Sommerfeld expansion is valid,
the entropy current can be directly related to the chemical-potential derivatives of the charge current~\cite{sup}
\begin{align}
j_a^s
=
-\frac{\pi^2T}{3}\partial_\mu j_a^e
-
\frac{\pi^4T^3}{45}\partial_\mu^3 j_a^e
+\cdots .
\end{align}
This relation also applies to both AC and DC responses and is not restricted to the linear-response regime.
In the linear DC limit,
its leading term reduces to the usual Mott-type relation~\cite{XCPRL2020} between the intrinsic anomalous charge and entropy Hall conductivities,
$\sigma_{ab}^s(T,\mu)=-(\pi^2T/3)\partial_\mu\sigma_{ab}^e(\mu)$,
where $\sigma_{ab}^e(\mu)$ is the zero-temperature intrinsic anomalous Hall conductivity.

\bigskip
\noindent\textbf{DISCUSSION}\\
In summary, we have developed a microscopic quantum theory of entropy transport for Bloch electrons,
starting from the single-particle entropy operator and its continuity equation.
Within the relaxation-time approximation,
the resulting entropy production reproduces the macroscopic Joule-heating relation order by order in the electric field,
while the positive leading contribution ensures consistency with the second law in the weak-field regime.
Microscopically, this establishes a direct link between dissipation and quantum geometry:
near topological Dirac band crossings and anticrossings,
the leading entropy production associated with the linear Drude current is controlled by the quantum metric,
whereas the subleading contribution arising from the nonlinear Drude
and intrinsic nonlinear Ohmic currents is controlled by the fully symmetrized quantum-metric dipole.
In contrast, Berry-curvature-induced linear and nonlinear Hall currents,
as well as the Hall component of the quantum-metric-dipole nonlinear current,
do not contribute to entropy production.
These results provide a microscopic quantum-geometric indicator for distinguishing dissipative from dissipationless transport.

We further predict a Berry-curvature-induced entropy Hall effect,
placing entropy transport on the same footing as charge and heat transport in the first law of thermodynamics.
A distinctive advantage of the entropy-current formulation is that,
for the steady and uniform response considered here,
its local definition directly yields the transport entropy current.
Consequently, in the near-equilibrium regime the transport heat current follows from $j_a^q=Tj_a^s$
without requiring an independent evaluation of the orbital magnetization.
Conversely, combining the transport entropy Hall current with the local heat Hall current isolates the magnetization current through
$j_a^m=j_a^h-Tj_a^s$ and reproduces the modern theory of orbital magnetization.

Beyond the near-equilibrium linear regime, heat and entropy currents are generally distinct,
and $j_a^q=Tj_a^s$ no longer provides a universal relation.
We therefore establish Maxwell-type and Sommerfeld relations between entropy and charge currents,
which apply beyond the linear DC limit and provide practical routes to infer entropy flow from charge-transport measurements.
{Our formulation is currently restricted to noninteracting Bloch electrons,
for which the entropy operator is constructed from the single-particle density matrix.
Extending the present framework to interacting systems and determining the underlying quantum geometric structures of entropy production
constitute important directions for future work.}

\bigskip
\noindent \textbf{\large METHODS}\\
\noindent\textbf{Quantum-metric dipole nonlinear current}---{For the intrinsic nonlinear conductivity
generated by the quantum metric dipole, both semiclassical and response theories give~\cite{sup}
\begin{align}
\sigma_{abc}
=
\sum_{m,n\neq m}\int_k
\frac{
v_m^a g^{bc}_{mn}
-
v_m^b g^{ac}_{mn}
-
v_m^c g^{ab}_{mn}
}
{\epsilon_{mn}}
f'_m.
\end{align}
Its fully symmetric component is
\begin{align}
\sigma^{O}_{abc}
=
-\frac{1}{3}
\sum_{mn}^{m \neq n} \int_k
\dfrac{
g^{abc}_{mn}
}
{\epsilon_{mn}}
f'_m,
\end{align}
where $g^{abc}_{mn} = v_m^a g^{bc}_{mn} + v_m^b g^{ca}_{mn} + v_m^c g^{ab}_{mn} $.
This component gives the dissipative intrinsic nonlinear Ohmic current $j_a^O=\sigma_{abc}^OE_bE_c$
and therefore appears in the subleading entropy production discussed in the main text.
The remaining Hall component is
\begin{align}
\sigma^{H}_{abc}
=
\frac{2}{3}
\sum_{mn}^{m \neq n} \int_k
\dfrac{\bar{g}^{abc}_{mn}}
{\epsilon_{mn}}
f'_m,
\end{align}
where $\bar{g}^{abc}_{mn}= 2v_m^a g^{bc}_{mn} - v_m^b g^{ca}_{mn} - v_m^c g^{ab}_{mn} $.
Although this component is also induced by the quantum-metric dipole,
it does not appear in the subleading entropy production and therefore represents a dissipationless nonlinear Hall current.
For a system with $C_{3z}$ symmetry, the genuine Hall part of the in-plane dc second-order conductivity,
$\sigma_{abc}^{H}$, vanishes, while only the Ohmic part $\sigma_{abc}^{O}$ can remain nonzero.
Its off-diagonal components can nevertheless generate a transverse current for particular electric-field directions~\cite{Wang2023}.}

{We remark that the factor $2/3$ in \(\sigma^{H}_{abc}\) has a transparent semiclassical explanation.
It results from the combination of the field-induced anomalous velocity $\vect{E}\times\vect{\Omega}_n^{E}$ and the field-induced group velocity.
The anomalous-velocity contribution is~\cite{GaoY2014PRL, BPT1, BPT2}
$j^{\text{I}}_a
=
-\sum_m\int_k f_m
(\vect{E} \times \boldsymbol{\Omega}^{E}_m)_a
=
\sum_m\int_k
\left(
G^{bc}_m\partial_a f_m
-
G^{ac}_m\partial_b f_m
\right)E_bE_c$,
where $G^{bc}_m=2\sum_{n \neq m}g^{bc}_{mn}/\epsilon_{mn}$
is the Berry-connection polarizability. After symmetrizing the two electric-field indices $b$ and $c$,
because the physical dc response is contracted with $E_bE_c=E_cE_b$, this contribution becomes
\begin{align}
j^{\text{I}}_a
=
\sum_{mn}^{m \neq n}\int_k
\frac{
2v_m^a g^{bc}_{mn}
-
v_m^b g^{ca}_{mn}
-
v_m^c g^{ab}_{mn}
}
{\epsilon_{mn}}
f'_m E_bE_c .
\end{align}
The field-induced correction to the band energy,
$\epsilon_m^{(2)}
=
-G^{bc}_mE_bE_c/2
=
-\sum_{n\neq m}
g^{bc}_{mn}E_bE_c/\epsilon_{mn} ,
$
generates a group-velocity contribution~\cite{YanPRL2024, Jia2024}
$j^{\text{II}}_a
=
-\sum_m\int_k f_m\partial_a\epsilon_m^{(2)}
=
-\sum_{mn}^{m \neq n}\int_k
\frac{
v_m^a g^{bc}_{mn}
}
{\epsilon_{mn}}
f'_m E_bE_c$.
This term can be decomposed into an Ohmic part and a Hall part as
$j^{\text{II}}_a=j^{\text{II;O}}_a+j^{\text{II;H}}_a$, where
\begin{align}
j^{\text{II;O}}_a
&=
-\frac{1}{3}
\sum_{m,n\neq m}\int_k
\frac{ g^{abc}_{mn} } {\epsilon_{mn}}
f'_m E_bE_c,
\\
j^{\text{II;H}}_a
&=
-\frac{1}{3}
\sum_{m,n\neq m}\int_k
\frac{ \bar{g}^{abc}_{mn} }{\epsilon_{mn}}
f'_m E_bE_c .
\end{align}
As a result, $j^{\text{II;O}}_a=\sigma_{abc}^OE_bE_c$ gives the intrinsic nonlinear Ohmic current
with the fully symmetric conductivity $\sigma^{O}_{abc}$ determined by $g^{abc}_{mn}$,
while the combination $j^{\text{I}}_a+j^{\text{II;H}}_a=\sigma_{abc}^HE_bE_c$ gives the nonlinear Hall current with the conductivity $\sigma^{H}_{abc}$
determined by $\bar{g}^{abc}_{mn}$.
}

{
The nonlinear Drude conductivity is~\cite{sup}
$\sigma_{abc}^{D'} = -\tau^2/2 \sum_m\int_{k} v_m^a\partial_b\partial_c f_m$,
which is fully symmetric under permutations of $a$, $b$, and $c$, and hence can be rewritten as
$\sigma_{abc}^{D'} 
=
\tau^2/2\sum_m \int_k f_m' v_m^c\partial_bv_m^a
=
\tau^2/6\sum_m\int_{k}f_m'
\left(
v_m^c\partial_bv_m^a
+
v_m^a\partial_cv_m^b
+
v_m^b\partial_av_m^c
\right)$.
Using $\partial_bv_m^a = w_{mm}^{ab} + 2\sum_{n \neq m}\epsilon_{mn}g_{mn}^{ab}$~\cite{XiangMap},
neglecting the nongeometric contribution $w^{ab}_{mm}$ near topological Dirac band crossings and anticrossings that we focus on,
we obtain
\begin{equation}
\sigma_{abc}^{D'}
=
\frac{\tau^2}{3}
\sum_{mn} \int_{k}
\epsilon_{mn}g_{mn}^{abc}f_m'.
\end{equation}
}

\noindent\textbf{Relation between optical charge and entropy currents}---{For interband optical responses,
the charge and entropy conductivities are given by~\cite{sup}
\begin{align}
\sigma_{ab}^e(\omega)
&=
\pi\omega\sum_{nm}\int_k
r_{nm}^a r_{mn}^b f_{nm}\delta(\epsilon_{mn}-\omega),
\\
\sigma_{ab}^s(\omega)
&=
\pi\omega\sum_{nm}\int_k
r_{nm}^a r_{mn}^b \Delta^s_{mn}\delta(\epsilon_{mn}-\omega).
\end{align}
For a two-band model $\hat{\mathcal{H}}=\vect{d}\cdot\hat{\vect{\sigma}}$,
these two conductivities are related by~\cite{sup}
\begin{align}
\frac{\sigma_{ab}^e(\omega)}{\sigma_{ab}^s(\omega)}
=
\frac{f_{-+}(\omega)}{\Delta^s_{+-}(\omega)},
\end{align}
where $\Delta^s_{+-}=s_+-s_-$ and $f_{-+}=f_--f_+$.
For the noninteracting Bloch electrons,
the entropy weights $s_n=-f_n\ln f_n-(1-f_n)\ln(1-f_n)$ are fixed by the band energies, chemical potential, and temperature.
These relations show that entropy currents can be inferred from the temperature,
chemical-potential, and frequency dependence of charge-current measurements.
Valley-selective optical selection rules~\cite{valley, Li2026},
when present, could further restrict the resonant transitions to a small region of momentum space
and thereby simplify the experimental interpretation of this relation.}

\bigskip
\noindent \textbf{\large DATA AVAILABILITY}\\
No external datasets were used in this study.

\bigskip
\noindent \textbf{\large CODE AVAILABILITY}\\
The numerical results in Figs. 2 and 3 are generated directly from the theoretical models.
The Python scripts used to reproduce these results and figures are available at Zenodo under DOI: \href{https://doi.org/10.5281/zenodo.23210799}{10.5281/zenodo.23210799}.

\newpage
\bigskip
\def\bibsection{\ }

\bigskip
\noindent \textbf{\large FUNDING STATEMENT}\\
This work was supported by the National Natural Science Foundation of China (Grants No. 12674063 and No. 12404059).

\bigskip
\noindent \textbf{\large AUTHOR CONTRIBUTIONS}\\
J.W. conceived the project.
L.J.X. and J.W. developed the theory.
L.J.X. and J.W. wrote the paper.
J.W. supervised the project.
All authors analyzed the data and contributed to the discussion of the results.

\bigskip
\noindent \textbf{\large COMPETING INTERESTS}\\
The authors declare no competing interests.

\bigskip
\noindent \textbf{\large FIGURE CAPTIONS}\\
\noindent\textbf{Fig. 1.} Quantum geometric entropy transport in crystalline solids.
The single-particle entropy operator leads to quantum-mechanical expressions for entropy production and entropy current through the entropy continuity equation.
Within the relaxation-time approximation and near topological Dirac band crossings and anticrossings,
the leading and subleading contributions to entropy production are controlled by the quantum metric $g_{mn}^{ab}$ and the fully symmetrized quantum-metric dipole $g_{mn}^{abc}$,
respectively. In addition to the anomalous Hall effect (HE) and the anomalous Ettingshausen effect (EE),
Berry curvature $\bm{\Omega}_n$ also gives rise to an entropy HE.

\noindent\textbf{Fig. 2.} Entropy production in the tilted Dirac model
$\hat{\mathcal{H}}_0=t_xk_x+v_xk_x\hat{\sigma}_x+v_yk_y\hat{\sigma}_y+\Delta\hat{\sigma}_z$.
(a) Leading entropy production for $\tau=10^{-14}$ s. (b) and (c) Nonlinear Drude contribution $\Pi_D^{(3)}$,
intrinsic nonlinear Ohmic contribution $\Pi_O^{(3)}$, and their sum $\Pi^{(3)}=\Pi_D^{(3)}+\Pi_O^{(3)}$ for $\tau=10^{-14}$ s and $10^{-15}$ s, respectively.
Parameters are $t_x=0.2\,{\rm eV}\cdot\text{\AA}$, $v_x=v_y=1\,{\rm eV}\cdot\text{\AA}$, $\Delta=0.1$ eV, $T=50$ K, and $E_x=10^4$ V/m.
The vertical dashed lines mark the band edges.

\noindent\textbf{Fig. 3.} Entropy Hall effect in the ferromagnetic Rashba model~\cite{Xiao2010}
$\hat{\mathcal{H}}_0=k^2/2m+\lambda_R(k_y\hat{\sigma}_x-k_x\hat{\sigma}_y)+\Delta\hat{\sigma}_z$, where $k^2=k_x^2+k_y^2$.
(a) Band dispersions, where the vertical dashed lines mark the band edges.
(b) Normalized Berry curvature of the lower band.
(c) Entropy weight $s_n$ as a function of $\epsilon_n-\mu$ at different temperatures, peaked at $\epsilon_n=\mu$ with $s_n^{\rm max}=\ln 2$.
(d) Entropy Hall conductivity $\sigma_{xy}^s$ as a function of $\mu$ at different temperatures.
Parameters are $\lambda_R=0.6\,{\rm eV}\cdot\text{\AA}$, $m=1.1m_e$, and $\Delta=0.1$ eV.

\widetext
\clearpage

\end{document}